\documentclass[zpreprint,zbstdefault]{zeus_paper}
\usepackage[english]{babel}
\usepackage{verbatim}
\usepackage{psfrag}
\newcommand{\ZcoosysB}{%
The ZEUS coordinate system is a right-handed Cartesian system, with the $Z$
axis pointing in the proton beam direction, referred to as the ``forward
direction'', and the $X$ axis pointing left towards the centre of HERA.
The coordinate origin is at the nominal interaction point.\xspace}

\chardef\usc=95
\chardef\til=126
\catcode`\@=11 
\DeclareRobustCommand\xdotspace{\futurelet\@let@token\@xdotspace}
\def\@xdotspace{%
  \ifx\@let@token.\else
  \ifx\@let@token\bgroup.\else
  \ifx\@let@token\egroup.\else
  \ifx\@let@token\/.\else
  \ifx\@let@token\ .\else
  \ifx\@let@token~.\else
  \ifx\@let@token!.\else
  \ifx\@let@token,.\else
  \ifx\@let@token:.\else
  \ifx\@let@token;.\else
  \ifx\@let@token?.\else
  \ifx\@let@token/.\else
  \ifx\@let@token'.\else
  \ifx\@let@token).\else
  \ifx\@let@token-.\else
  \ifx\@let@token\@xobeysp.\else
  \ifx\@let@token\space.\else
  \ifx\@let@token\@sptoken.\else
   .\space
   \fi\fi\fi\fi\fi\fi\fi\fi\fi\fi\fi\fi\fi\fi\fi\fi\fi\fi}
\catcode`\@=12 

\newcommand{\stru}[2]{%
   \relax\ifmmode\hbox{\vrule height#1 depth#2 width0pt}%
   \else\vrule height#1 depth#2 width0pt\fi}

\newcommand{\Ronum}[1]{\uppercase\expandafter{\romannumeral#1}}
\newcommand{\ronum}[1]{\expandafter{\romannumeral#1}}
\DeclareRobustCommand{\LaTeXZ}{%
  \LaTeX\kern-.05em4\kern-.1em
  {\raisebox{-0.2ex}{$\scriptstyle\text{ZEUS}$}}\xspace}

\DeclareMathAlphabet{\mathbf}{OT1}{cmr}{bx}{sl}
\newcommand{\eVdist}{\kern-0.06667em}

\newcommand{\Gev}{{\text{Ge}\eVdist\text{V\/}}}

\newcommand{\mev}{{\,\text{Me}\eVdist\text{V\/}}}
\newcommand{\gev}{{\,\text{Ge}\eVdist\text{V\/}}}

\newcommand{\pbi}{\,\text{pb}^{-1}}

\newcommand{\cm}{\,\text{cm}}

\newcommand{\Tesla}{\,\text{T}}

\newcommand{\slashfrac}[2]{%
  \raisebox{0.5ex}{\ensuremath #1}\kern-0.12em/\kern-0.08em
  \raisebox{-.8ex}{\ensuremath #2}}

\newcommand{\sqr}[3]{%
    {\vcenter{\hrule height.#3ex\hbox{\vrule width.#2ex height#1ex
     \kern#1ex\vrule width.#3ex}\hrule height.#2ex}}}

\newcommand{\widebar}[1]{%
   \mkern1.5mu\overline{\mkern-1.5mu#1\mkern-1.mu}\mkern1.mu}
\catcode`\@=11 
\newcommand{\parenbar}{\mathpalette\p@renb@r}
\def\p@renb@r#1#2{\vbox{%
  \ifx#1\scriptscriptstyle \dimen@.7em\dimen@ii.2em\else
  \ifx#1\scriptstyle \dimen@.8em\dimen@ii.25em\else
  \dimen@1em\dimen@ii.4em\fi\fi \offinterlineskip
  \ialign{\hfill##\hfill\cr
    \vbox{\hrule width\dimen@ii}\cr
    \noalign{\vskip-.3ex}%
    \hbox to\dimen@{$\mathchar300\hfil\mathchar301$}\cr
    \noalign{\vskip-.3ex}%
    $#1#2$\cr}}}
\catcode`\@=12 

\newcommand{\rnge}{\hbox{$\,\text{--}\,$}}

\newcommand{\IP}{{\rm I$\kern-0.01667em$P}\xspace}

\newcommand{\Lumi}{{\cal L}}

\mathchardef\qsm=63
\mathchardef\pls=43
\mathchardef\mns=512
\mathchardef\plm=518
\mathchardef\eql=61
\mathchardef\smallleft=300
\mathchardef\smallright=301
\mathchardef\les=316
\mathchardef\gre=318
\mathchardef\leq=532
\mathchardef\grq=533

\catcode`\@=11 
\newcounter{pict@width}
\newcounter{pict@height}
\newlength{\pict@scale}
\newcommand{\psfigadd}[4]{%
\setcounter{pict@width}{1*\ratio{#2+\pict@scale/2}{\pict@scale}}
\setcounter{pict@height}{1*\ratio{#3+\pict@scale/2}{\pict@scale}}
\setlength{\unitlength}{\pict@scale}
\hbox to #2{\hspace{-\fill}\begin{picture}(\thepict@width,\thepict@height)
\put(0,0){\psfig{figure=#1,width=#2,height=#3,clip=}}
\SetScale{0.283466457}
\SetWidth{1.763889}
{#4}
\end{picture}}
}
\newcounter{pict@widthfst}
\newcounter{pict@widthscd}
\newcounter{pict@widthtot}
\newcommand{\psfigaddtwo}[7]{%
\setcounter{pict@widthfst}{1*\ratio{#2+\pict@scale/2}{\pict@scale}}
\setcounter{pict@widthscd}{1*\ratio{#2+#4+\pict@scale/2}{\pict@scale}}
\setcounter{pict@widthtot}{1*\ratio{#2+#4+#6+\pict@scale/2}{\pict@scale}}
\setcounter{pict@height}{1*\ratio{#3+\pict@scale/2}{\pict@scale}}
\setlength{\unitlength}{\pict@scale}
\hbox{\hspace{-\fill}\begin{picture}(\thepict@widthtot,\thepict@height)
\put(0,0){\psfig{figure=#1,width=#2,height=#3,clip=}}
\put(\thepict@widthscd,0){\psfig{figure=#5,width=#6,height=#3,clip=}}
\SetScale{0.283466457}
\SetWidth{1.763889}
{#7}
\end{picture}}
}
\newcommand{\psfigror}[4]{%
\setcounter{pict@width}{1*\ratio{#2+\pict@scale/2}{\pict@scale}}
\setcounter{pict@height}{1*\ratio{#3+\pict@scale/2}{\pict@scale}}
\setlength{\unitlength}{\pict@scale}
\hbox{\begin{picture}(\thepict@width,\thepict@height)
\put(0,\thepict@height){\psfig{figure=#1,width=#3,height=#2,clip=,angle=270}}
\SetScale{0.283466457}
\SetWidth{1.763889}
{#4}
\end{picture}}
}
\newcommand{\psfigrol}[4]{%
\setcounter{pict@width}{1*\ratio{#2+\pict@scale/2}{\pict@scale}}
\setcounter{pict@height}{1*\ratio{#3+\pict@scale/2}{\pict@scale}}
\setlength{\unitlength}{\pict@scale}
\hbox{\begin{picture}(\thepict@width,\thepict@height)
\put(0,0){\psfig{figure=#1,width=#3,height=#2,clip=,angle=90}}
\SetScale{0.283466457}
\SetWidth{1.763889}
{#4}
\end{picture}}
}
\catcode`\@=12 
\newlength\listtextwidth

\catcode`\@=11 
\newlength{\@tabfninsert}
\newlength{\@tabfnwidth}
\newcommand{\tabfootnote}[2]{%
  \setlength{\@tabfninsert}{0.8em}
  \setlength{\@tabfnwidth}{\textwidth}
  \addtolength{\@tabfnwidth}{-\@tabfninsert}
  \addtolength{\@tabfnwidth}{-0.4em}
  \noindent\makebox[\@tabfninsert][r]{\footnotesize$^{#1}$\hfil}\hfill%
  \parbox[t]{\@tabfnwidth}{\footnotesize #2\hfill}}
\catcode`\@=12 

\newcommand{\by}[2]{\leavevmode\kern.1em
\raise.5ex\hbox{\ensuremath{\scriptstyle #1}}\kern-.1em
/\kern-.15em\lower.25ex\hbox{\ensuremath{\scriptstyle #2}}}

\def\citeCTD{{\cite{%
nim:a279:290,*npps:b32:181,*nim:a338:254%
}}\xspace}
\def\citeCAL{{\cite{%
nim:a309:77,*nim:a309:101,*nim:a321:356,*nim:a336:23%
}}\xspace}

\begin{document}

\prepnum{DESY--05--071}

\title{
      Measurement of inelastic \(J/\psi\) production in deep inelastic
                            scattering at HERA \\
}

\author{ZEUS Collaboration}
\draftversion{}
\date{\today}

\abstract{
The inelastic production of \(J/\psi\) mesons in \(e p\) collisions has
been studied with the ZEUS detector at HERA using an integrated luminosity of
\(109\pbi\). The \(J/\psi\) mesons were identified using the decay channel
\(J/\psi\to\mu^{+}\mu^{-}\). The measurements were performed in the kinematic 
range \(2<Q^2<80\gev^2\), \(50<W<250\gev\), \(0.2<z<0.9\) and
\(-1.6<Y_\text{lab}<1.3\), where \(Q^2\) is the virtuality of the
exchanged photon, \(W\) is the photon--proton centre--of--mass energy, 
\(z\) is the fraction of the photon
energy carried by the \(J/\psi\) meson in the proton rest frame and 
\(Y_\text{lab}\) is the rapidity of the \(J/\psi\) in the laboratory frame.
The measured cross sections are compared to theoretical 
predictions within the non-relativistic QCD framework including 
colour--singlet and colour--octet contributions, as well as to predictions based on 
the \(k_T\)--factorisation approach.
Calculations of the colour--singlet process generally agree with the 
data, whereas inclusion of colour--octet terms spoils this agreement.
}

\makezeustitle

\def\3{\ss}                                                                                        
\newcommand{\address}{ }                                                                           
\pagenumbering{Roman}                                                                              
                                                   %
\begin{center}                                                                                     
{                      \Large  The ZEUS Collaboration              }                               
\end{center}                                                                                       
  S.~Chekanov,                                                                                     
  M.~Derrick,                                                                                      
  S.~Magill,                                                                                       
  S.~Miglioranzi$^{   1}$,                                                                         
  B.~Musgrave,                                                                                     
  \mbox{J.~Repond},                                                                                
  R.~Yoshida\\                                                                                     
 {\it Argonne National Laboratory, Argonne, Illinois 60439-4815}, USA~$^{n}$                       
\par \filbreak                                                                                     
  M.C.K.~Mattingly \\                                                                              
 {\it Andrews University, Berrien Springs, Michigan 49104-0380}, USA                               
\par \filbreak                                                                                     
  N.~Pavel, A.G.~Yag\"ues Molina \\                                                                
  {\it Institut f\"ur Physik der Humboldt-Universit\"at zu Berlin,                                 
           Berlin, Germany}
% ...
\newpage                                                                                           
$^{\    1}$ also affiliated with University College London, UK \\                                  
% ...
                                                           %
                                                           %
\newpage   
                                                           %
                                                           %
\begin{tabular}[h]{rp{14cm}}                                                                       
$^{a}$ &  supported by the Natural Sciences and Engineering Research Council of Canada (NSERC) \\  
$^{b}$ &  supported by the German Federal Ministry for Education and Research (BMBF), under        
          contract numbers HZ1GUA 2, HZ1GUB 0, HZ1PDA 5, HZ1VFA 5\\                                
$^{c}$ &  supported in part by the MINERVA Gesellschaft f\"ur Forschung GmbH, the Israel Science   
          Foundation (grant no. 293/02-11.2), the U.S.-Israel Binational Science Foundation and    
          the Benozyio Center for High Energy Physics\\                                            
$^{d}$ &  supported by the German-Israeli Foundation and the Israel Science Foundation\\           
$^{e}$ &  supported by the Italian National Institute for Nuclear Physics (INFN) \\                
$^{f}$ &  supported by the Japanese Ministry of Education, Culture, Sports, Science and Technology 
          (MEXT) and its grants for Scientific Research\\                                          
$^{g}$ &  supported by the Korean Ministry of Education and Korea Science and Engineering          
          Foundation\\                                                                             
$^{h}$ &  supported by the Netherlands Foundation for Research on Matter (FOM)\\                   
$^{i}$ &  supported by the Polish State Committee for Scientific Research, grant no.               
          620/E-77/SPB/DESY/P-03/DZ 117/2003-2005 and grant no. 1P03B07427/2004-2006\\             
$^{j}$ &  partially supported by the German Federal Ministry for Education and Research (BMBF)\\   
$^{k}$ &  supported by RF Presidential grant N 1685.2003.2 for the leading scientific schools and  
          by the Russian Ministry of Education and Science through its grant for Scientific        
          Research on High Energy Physics\\                                                        
$^{l}$ &  supported by the Spanish Ministry of Education and Science through funds provided by     
          CICYT\\                                                                                  
$^{m}$ &  supported by the Particle Physics and Astronomy Research Council, UK\\                   
$^{n}$ &  supported by the US Department of Energy\\                                               
$^{o}$ &  supported by the US National Science Foundation\\                                        
$^{p}$ &  supported by the Polish Ministry of Scientific Research and Information Technology,      
          grant no. 112/E-356/SPUB/DESY/P-03/DZ 116/2003-2005 and 1 P03B 065 27\\                  
$^{q}$ &  supported by FNRS and its associated funds (IISN and FRIA) and by an Inter-University    
          Attraction Poles Programme subsidised by the Belgian Federal Science Policy Office\\     
\end{tabular}                                                                                      
                                                           %
                                                           %

\pagenumbering{arabic} 
\pagestyle{plain}
\section{Introduction}
\label{sec-int}

Inelastic production of charmonium can be described in two steps.
The first step is the creation of a \(c\widebar{c}\) quark pair, a process which 
can be calculated in perturbative Quantum Chromodynamics (QCD). 
The second step is the formation of the \(J/\psi\) bound state, 
which occurs at long distances and is described by phenomenological 
models. 

When charmonium production was first investigated at 
CERN~\cite{pl:b258:493,*np:b213:1,*pl:b256:112} and 
Fermilab~\cite{prl:69:3704}
it was hoped that the production cross section could be used to
determine the gluon density in the proton, because the gluon density directly 
enters the cross--section calculation. This idea was encouraged by the 
qualitative agreement of the data with the predictions of LO QCD within
the framework of the colour--singlet
model (CSM)~\cite{pr:d23:1521} in which the \(c\widebar{c}\) pair is formed 
in a CS state identified with the \(J/\psi\). Later data 
from \(p\widebar{p}\) collisions at Fermilab~\cite{prl:79:572,*prl:79:578} 
indicated that the CSM is not able to describe \(J/\psi\) production at
large \(J/\psi\) transverse momenta, and hence that there may be significant
contributions from higher orders in QCD or from the production of 
\(c\widebar{c}\) pairs in colour--octet (CO) states, which evolve 
into \(J/\psi\) mesons via radiation of soft gluons.

Models have been developed in the framework of non-relativistic QCD 
(NRQCD)~\cite{pr:d51:1125} in which CS and CO contributions coexist. The 
transition of the coloured \(c\widebar{c}\) pair, with a given angular momentum,  
into a \(J/\psi\) is described in terms of long--distance matrix elements 
tuned to experimental data (hadroproduction of \(J/\psi\) mesons or 
\(B\)--meson decays to \(J/\psi\)).
As well as describing high--\(p_T\) charmonium production in \(p\widebar{p}\) 
collisions~\cite{prl:79:572,*prl:79:578}, NRQCD
calculations including CS and CO contributions are also consistent with the 
data on production of \(J/\psi\) mesons in \(\gamma\gamma\) interactions at 
LEP2~\cite{pl:b565:76}. However, \(J/\psi\) polarisation data from 
CDF~\cite{prl:85:2886} are inconsistent with NRQCD predictions. 
Comparisons with the decay angular distributions measured in \(e^+ e^-\) 
collisions at BaBar~\cite{prl:87:162002} and Belle~\cite{prl:88:52001} are 
inconclusive. 

The production of \(J/\psi\) mesons in $ep$ collisions at HERA is sensitive 
to both CS and CO contributions.
The CS mechanism is expected to be the dominant contribution at
intermediate values of the inelasticity variable, \(z \lesssim 0.7 \).
In the proton rest frame, \(z\) is the fraction of the virtual photon 
energy transferred to the \(J/\psi\).
The CO mechanism is expected to be dominant at high \(z\).
However, large contributions from the elastic and diffractive 
proton--dissociative \(J/\psi\) production 
processes~\cite{pl:b483:23,*pl:b568:205,epj:c24:345,*epj:c26:389} 
are also present at $z \approx 1$.

Inelastic \(J/\psi\) production at HERA was studied previously in the
photoproduction regime (photon virtuality \(Q^2 \approx 0\)) by the 
H1~\cite{epj:c25:25} and ZEUS~\cite{epj:c27:173} collaborations.
The leading--order (LO) NRQCD calculations
and the next--to--leading-order (NLO) CSM predictions are 
both consistent with the data.  Inelastic \(J/\psi\) production
in the deep inelastic scattering (DIS) regime (\(Q^2 \gtrsim 1 \gev^2\))
has been studied by the H1 collaboration~\cite{epj:c25:41}.
In this case, the LO NRQCD predictions overestimate the data, 
whereas the LO CSM expectations underestimate them.  The
shape of the differential cross sections are reasonably
well reproduced by both models, except for the inelasticity distribution 
in the case of LO NRQCD and for the distribution of the \(J/\psi\) 
transverse--momentum squared
in the photon--proton centre--of--mass system in the case of the LO CSM.
NLO CSM predictions are not available in the DIS regime.

Models in the framework of the semi--hard or \(k_T\)--factorisation 
approach~\cite{prep:100:1,*prep:189:267,*sovjnp:53:657,*sovjnp:54:867,*pl:b242:97,*np:b366:135,*np:b360:3,*np:b386:215} are also available. 
In these models, based on non--collinear parton dynamics governed by the 
BFKL~\cite{jetp:44:443,*jetp:45:199,*sovjnp:28:822} or
CCFM~\cite{np:b296:49,*pl:b234:339} evolution equations,
effects of non--zero gluon transverse momentum are taken into account. 
Cross sections are then calculated as the convolution of unintegrated 
(transverse--momentum dependent) gluon densities and LO off--shell matrix 
elements.
These models~\cite{epj:c27:87,jp:g29:1395,pr:d66:114003} succeed in 
describing the \(p_T\) spectra of different quarkonium states at Fermilab and 
\(J/\psi\) meson production at HERA, as well as the quarkonium polarisation
properties measured both at Fermilab and HERA.

This paper presents a measurement of inelastic \(J/\psi\) production in 
DIS and comparison of the data with NRQCD and models in the \(k_T\)--factorisation 
approach. The reaction \(e\,p\to e\,J/\psi\,X\) 
is studied for \(Q^2>2\gev^2\). 
The photon--gluon fusion process leading to a \(J/\psi\) in the final
state is assumed to be the dominant mechanism. Other contributions arise from the
production of \(\psi'\) mesons which subsequently decay to a \(J/\psi\), and from 
\(J/\psi\) and \(\psi'\) production from the resolved photon process, where the 
exchanged photon acts as a source of partons; the latter process, however, is 
suppressed at high $Q^2$. In addition, beauty production at high $Q^2$ with 
subsequent decay of a 
$B$ meson into a \(J/\psi\) also contributes to the measured cross section; 
this process is suppressed due to the small beauty cross section and
the small $B$ to \(J/\psi\) branching ratio.

Inelastic \(J/\psi\) production at large $Q^2$ has a smaller cross section 
than in photoproduction but presents several interesting 
aspects. 
The contribution from the CO model is expected to be more significant;
both the CO and the CS predictions should be more accurate due to the higher 
scale in the interaction. Also, backgrounds from diffractive 
processes are reduced at high $Q^2$.
The measurements presented here are in a larger kinematic range than
those previously published~\cite{epj:c25:41}.
A measurement of the hadronic final state, $X$, is presented for the
first time.

\section{Experimental set-up}

The data sample used in the analysis was collected with the ZEUS detector 
in the years 1996--2000 and corresponds to an integrated luminosity 
\(\Lumi=108.8\pm2.2\pbi\).
During the 1996--97 data taking, HERA operated with positrons of energy 
\(E_e=27.5\gev\) and protons of energy \(E_p=820\gev\), corresponding to 
a  centre-of-mass energy  \(\sqrt s=300\gev\)
(\(\Lumi_{300}=32.7\pm0.6\pbi\)).
In the years 1998--2000, HERA collided electrons or positrons with 
protons of energy  \(E_p=920\gev\),  corresponding to
\(\sqrt{s}=318\gev\) (\(\Lumi_{318}=76.1\pm1.6\pbi\)).
The cross sections presented here were corrected to \(\sqrt{s}=318\gev\) 
using the Monte Carlo (MC) simulation described in Section~\ref{sec-MCsim}. 

A detailed description of the ZEUS detector can be found
elsewhere~\cite{pl:b293:465,zeus:1993:bluebook}. 
Charged particles are tracked in the central tracking detector (CTD)~\citeCTD,
which operates in a magnetic field of $1.43\Tesla$ provided by a thin 
superconducting coil. The CTD consists of 72~cylindrical drift chamber 
layers, organised in 9~superlayers covering the 
polar--angle\footnote{\ZcoosysB} region 
\mbox{$15^\circ<\theta<164^\circ$}. The transverse--momentum resolution for
full--length tracks is $\sigma(p_T)/p_T=0.0058p_T\oplus0.0065\oplus0.0014/p_T$,
with $p_T$ in $\Gev$.
Energy deposits are measured in the
high--resolution uranium--scintillator calorimeter (CAL)~\citeCAL 
which
consists 
of three parts: the forward (FCAL), the barrel (BCAL) and the rear (RCAL)
calorimeters. Each part is subdivided transversely into towers and
longitudinally into one electromagnetic section (EMC) and either one (in RCAL)
or two (in BCAL and FCAL) hadronic sections (HAC). The smallest subdivision of
the calorimeter is called a cell.  The CAL energy resolutions, as measured under
test-beam conditions, are $\sigma(E)/E=0.18/\sqrt{E}$ for electrons and
$\sigma(E)/E=0.35/\sqrt{E}$ for hadrons with $E$ in $\Gev$.
The scattered electron\footnote{
Here and in the following, the term ``electron'' denotes generically both the 
electron ($e^-$) and the positron ($e^+$).} 
identification is performed by combining information
from the CAL, the small--angle rear tracking detector
(SRTD)~\cite{nim:a401:63,epj:c21:443} and the hadron--electron separator
(HES)~\cite{nim:a277:176}.
Muon identification is performed by finding tracks in the barrel and rear 
muon chambers (BMUON and RMUON)~\cite{nim:a333:342} or minimum--ionising 
energy deposits in the CAL, matched to CTD tracks.
The muon chambers
are placed inside and outside a magnetised iron yoke surrounding the CAL.
The barrel and rear inner muon chambers, used in this analysis, cover polar 
angles from 34$^\circ$ 
to 135$^\circ$ and from 135$^\circ$ to 171$^\circ$, respectively.

The luminosity was determined from the rate of the electron--proton
 bremsstrahlung process, 
\(e p \rightarrow e \gamma p\), where the photon was measured by a 
lead--scintillator calorimeter~\cite{desy-92-066,*zfp:c63:391,*acpp:b32:2025}
located at $Z=-107$~m.

\section{Event selection and reconstruction}

\subsection{Event selection}

A three--level trigger system was used to select events 
online~\cite{zeus:1993:bluebook,uproc:chep:1992:222}.
The first-- and second--level trigger selections were based on the
identification of a scattered electron in the CAL, as described in detail
elsewhere~\cite{epj:c21:443}.
The third--level trigger selection required both a scattered electron in the 
CAL and a track segment reconstructed in the barrel or rear inner muon 
chambers. 

Additional requirements were imposed in the offline selection
in order to suppress the
photoproduction background and select inelastic events with a \(J/\psi\)
candidate identified by the presence of a pair of oppositely charged muons.
In addition, the \(Z\) position of the reconstructed vertex
was required to lie within \(50\cm\) of the nominal interaction point.

\subsection{Reconstruction of DIS kinematic variables}

A scattered electron candidate, identified from the pattern of the energy
deposits in the CAL~\cite{nim:a365:508,*nim:a391:360}, was required. The 
electron position measurement of the CAL was improved using information from 
the SRTD and the HES.
To ensure full containment of the electromagnetic shower, the electron
impact position on the inner face of the rear calorimeter was required to 
lie outside the box \(|X|<13\cm\), \(|Y|<7\cm\). 
The energy of the scattered electron was required to be greater than 
\(10\gev\).

The photon virtuality, \(Q^2\), was reconstructed from the polar angle 
and energy of the scattered electron and was required to be in the 
range \(2<Q^2<80\gev^2\).
The Bjorken variable, \(y=(P\cdot q)/(P\cdot k)\), where
\(P\), \(q\) and \(k\) are the four--momenta of the incoming
proton, exchanged photon and incoming electron, respectively,
was reconstructed with the \(\Sigma\) method~\cite{nim:a361:197}.
Monte Carlo studies showed this method to be the most precise in the selected 
phase space region.
The photon--proton centre--of--mass energy, \(W\), calculated from 
\(W^2=ys-Q^2\), was restricted to the range \(50<W<250\gev\).

Conservation of energy, $E$, and longitudinal momentum, $p_Z$, require
\(\delta=\Sigma_i~(E_i-p_{Z,i})=2E_e=55\gev\), where the sum runs over all
the particles in the final state.
The experimentally reconstructed quantity 
\(\delta_{{\rm meas}}=\Sigma_i~(E_i-p_{Z,i})\) was calculated, where the
sum runs over all final--state energy--flow
objects~\cite{epj:c1:81,*thesis:briskin:1998}
(EFOs) which combine the information from calorimetry and tracking.
Only events with \(40<\delta_{{\rm meas}}<65\gev\) were kept.
This cut reduces background from photoproduction events, where the
scattered electron is not detected, and removes DIS events 
with large initial--state radiation, where the incoming electron
radiates a high--energy photon before the interaction and the photon
escapes detection in the rear beam hole.
To reduce background from photoproduction events further, the  
condition \(y_e<0.95\) was applied, where \(y_e\) indicates the
value of \(y\) reconstructed from the scattered electron energy
and polar angle.
In order to ensure an accurate reconstruction of the final state,
it was also required that the value of \(y\), obtained with the
Jacquet--Blondel method~\cite{proc:epfac:1979:391}, be larger than
0.02.

\subsection{\(J/\psi\) reconstruction}

The oppositely charged muons from the \(J/\psi\) decays were reconstructed 
in the CTD. 
Each track considered in the analysis was required to be fitted to
the event vertex, to reach at least the third superlayer of the
CTD and to have transverse momentum \(p_T>100\mev\); this guarantees
good reconstruction quality. At least one of the tracks from the \(J/\psi\) 
decay had to match a 
segment in the inner muon chambers and the other
had to match a CAL cluster with an energy deposit
consistent with the passage of a minimum ionising particle
(mip). To ensure high muon--identification efficiency and purity,
the track matched with the segment in the barrel (rear) inner muon 
chambers was required to have \(p_T>1.4\gev\) (\(p>1.8\gev\),
where \(p\) is the track momentum).
For the track matched to the mip cluster in the CAL, the cut \(p>1\gev\) 
was applied.
The muon identification and reconstruction efficiencies
were estimated separately for muons reconstructed in the BMUON, RMUON 
and CAL using independent samples of dimuon events.
The efficiency for tracks matched to the inner muon chambers
varies from 35\% for \(p_T \sim 1.4\gev\) to 60\% at high transverse
momentum for the barrel inner muon chambers and from 50\% for 
\(p \sim 1.8\gev\) to 65\% at high momentum for the rear inner muon 
chambers.
The efficiency for tracks matched to a mip in the CAL was 92\%.

The \(J/\psi\) rapidity in the laboratory frame, defined as
\(Y_\text{lab}=\by12\ln[(E_\psi+p_{Z,\psi})/(E_\psi-p_{Z,\psi})]\),
where \(E_\psi\) and \(p_{Z,\psi}\) are the energy and
longitudinal momentum of the \(J/\psi\) meson, was limited to 
the region \(-1.6<Y_\text{lab}<1.3\), where the acceptance is high.

The inelasticity of the \(J/\psi\) meson, \(z=(P\cdot p_\psi)/(P\cdot q)\), 
where $p_\psi$ is the four--momentum of the \(J/\psi\), was
reconstructed using the expression
\[
z = \frac{E_\psi-p_{Z,\psi}}{2 \, E_e \, y_{\Sigma}},
\]
where \( y_{\Sigma} = \sum_i^\text{had}(E_i-p_{Z,i}) / \delta_{\rm meas} \) 
and the sum in the numerator runs over all EFOs not associated with the 
scattered electron.
According to MC studies, the average resolution in \(z\) is 10\%.
The inelasticity was restricted to the range \(0.2<z<0.9\).
The lower \(z\) cut removes the region of high non-resonant background 
due to fake muons and the upper \(z\) cut removes elastic 
\(J/\psi\) events and suppresses diffractive \(J/\psi\) production
with dissociation of the proton.
In order to suppress further the latter background, the following cuts
were applied:
\begin{itemize}
\item 
the analysis was restricted to events with an energy deposit greater
than \(1\gev\) in a cone of \(35^\circ\) along the outgoing proton beam
direction (excluding calorimeter deposits due to the decay muons);
\item the event was required to have at least one track in addition 
to those associated with the two muons and with the scattered electron.
\end{itemize}

\begin{figure}
\unitlength1cm  \begin{picture}(15.5,20.)
\includegraphics{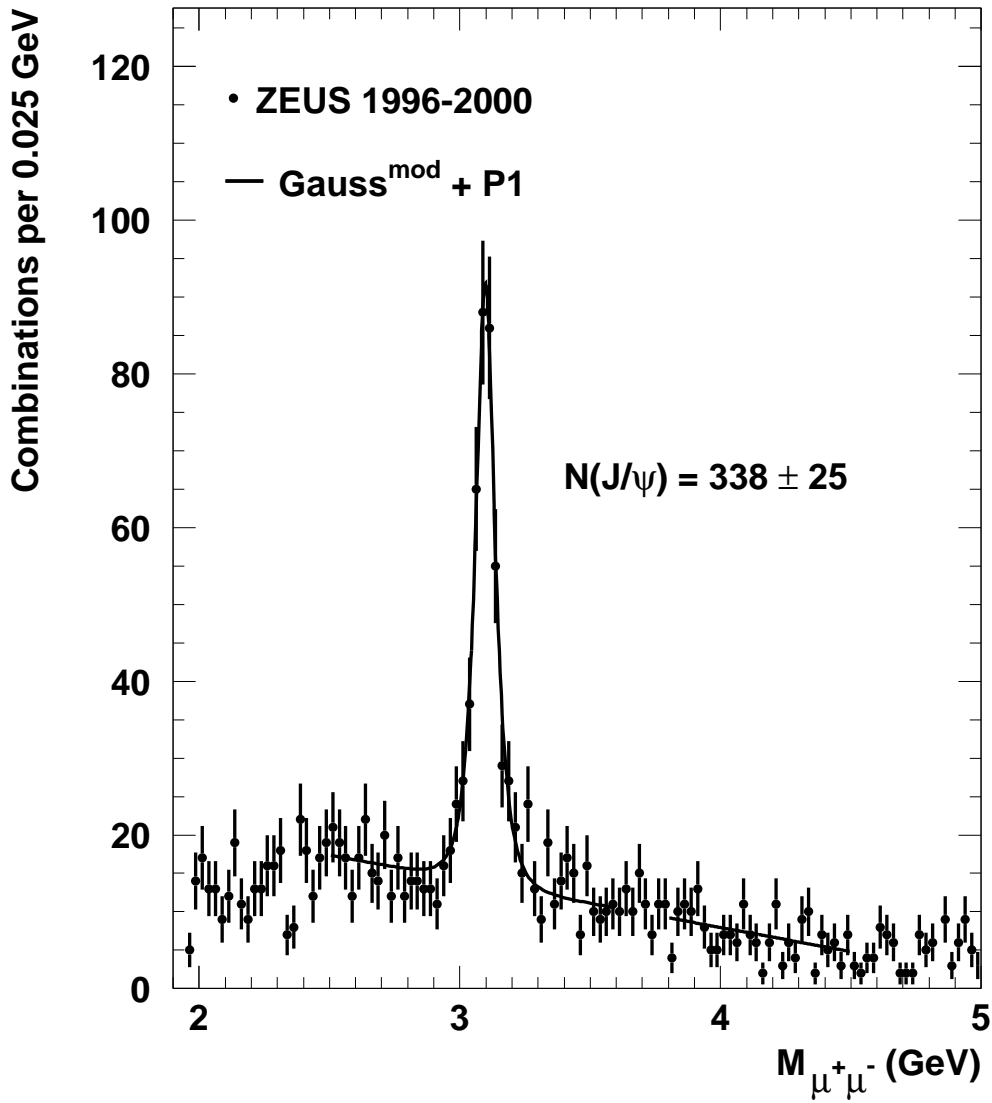}
\end{picture}
\vspace{-0.5cm}
\caption{
Invariant mass spectrum after all selection cuts in
the kinematic region \(2<Q^2<80\gev^2\), \(50<W<250\gev\), 
\(0.2<z<0.9\) and \(-1.6<Y_\text{lab}<1.3\). The curve is the
result of the fit with a modified Gaussian for the signal (see text) 
and a linear function (P1) for the non--resonant background.
}
\label{fig:mmumu}
\end{figure}

Figure~\ref{fig:mmumu} shows the invariant mass, \(M_{\mu^{+}\mu^{-}}\), 
distribution of all selected muon pairs. The distribution 
was fitted in the intervals \mbox{\(2.5<M_{\mu^{+}\mu^{-}}<3.6\gev\)}
and \mbox{\(3.8<M_{\mu^{+}\mu^{-}}<4.5\gev\)} with a function taken to be
the sum of a ``modified'' Gaussian, to describe the signal,
and a linear function, to describe the non-resonant background.
The range \(3.6<M_{\mu^{+}\mu^{-}}<3.8\gev\) was excluded to avoid
any overestimation of the background due to the \(\psi'\) state.
The modified Gaussian function had the form:
\[\text{Gauss}^\text{mod}\propto \exp [-0.5 \cdot x^{1+1/(1+0.5 \cdot x)}],\]
where \(x=|(M_{\mu^{+}\mu^{-}}-M_0)/\sigma|\). This function was
introduced to take into account the non--Gaussian tails of the
resonant signal.
This functional form describes both data and MC
signals well.
The position of the Gaussian, \(M_0\), the signal width, \(\sigma\),
as well as the number of signal events were free parameters of the fit. 
The fit yielded a peak position of \(M_0\) = 3098 $\pm$ 3 MeV, in 
agreement with the PDG value~\cite{pl:b592:1}, and a width of 
\mbox{\(\sigma\) = 35 $\pm$ 3 MeV}, in agreement with the MC estimation of
the detector resolution.
The number of \(J/\psi\) mesons was 338 $\pm$ 25.

\section{Monte Carlo models}
\label{sec-MCsim}

Inelastic \(J/\psi\) events were generated using the
{\sc Epjpsi}~\cite{proc:hera:1991:1488,*zfp:c60:721} MC generator. 
{\sc Epjpsi} incorporates the photon--gluon fusion process at LO, with
initial-- and final--state parton showers performed according to the
colour--dipole model as implemented in {\sc Ariadne}~\cite{cpc:71:15}.
\(J/\psi\) mesons were produced in the framework of the CSM.
The GRV98~\cite{epj:c5:461} parton distribution functions were used.
The scales for the evaluation of the strong coupling constant
and the proton structure function were set to the centre--of--mass
energy in the \(\gamma^{*} g\) frame.
The hadronisation was performed with the Lund string model~\cite{prep:97:31}.
The {\sc Epjpsi} MC predictions were reweighted to the data in \(Q^2\) and 
\(p^{* 2}_T\), where \(p^{* 2}_T\) is \(J/\psi\) transverse momentum squared
in the \(\gamma p\) centre--of--mass frame.
The helicity parameter in the {\sc Epjpsi} MC was
set to zero; this hypothesis is supported by the data~\cite{epj:c27:173}.

Signal events were also generated using the {\sc Cascade} MC
program~\cite{epj:c19:351}.
{\sc Cascade} incorporates the off-shell matrix elements for the 
photon--gluon fusion process at LO. 
The initial--state parton shower is generated according to the CCFM
evolution equations~\cite{np:b296:49,*pl:b234:339}.
The \(J/\psi\) mesons were produced in the framework of the CSM. 
The  gluon density, unintegrated in transverse momentum, $k_{T}$, was
obtained from an analysis of the proton structure functions based on
the CCFM equations~\cite{hep-ph-0309009}; in the event generation the gluon
density used corresponds to the set named {``J2003 set 2''}.
In {\sc Cascade} the hadronisation was also performed with the
Lund string model.

Events with diffractive dissociation of the proton, \(e p \to e J/\psi N\),
where \(N\) is a low mass state with the quantum numbers of the proton,
were simulated using the {\sc Epsoft} MC generator~\cite{thesis:kasprzak:1994}, 
which has been tuned to describe such processes at 
HERA~\cite{thesis:adamczyk:1999}.
Proton--dissociative events were also simulated with
the {\sc Diffvm}~\cite{diffvm} MC generator. {\sc Diffvm} has a more
detailed simulation of the final state than {\sc Epsoft}.

\(J/\psi\) mesons originating from \(B\)--meson decay were simulated
using the {\sc Rapgap} MC generator~\cite{cpc:86:147},
via the photon--gluon fusion process, \(\gamma^{*}g\to b\widebar{b}\);
the beauty--quark mass was set to 4.75 \(\gev\).
The CTEQ5L~\cite{epj:c12:375} parton distribution functions were used.
The \(B\) to \(J/\psi\) branching ratio in {\sc Rapgap} was set to the
PDG value~\cite{pl:b592:1}.
The MC prediction was normalised to the measured beauty cross
section in DIS~\cite{pl:b599:173}.

All generated events were passed through a full simulation of the ZEUS
detector based on {\sc Geant} 3.13~\cite{tech:cern-dd-ee-84-1}.
They were then subjected to the same trigger requirements and processed
by the same reconstruction programmes as for the data.

\section{Cross-section calculation}

Prior to the cross--section calculation, the residual diffractive
proton--dissociative background was subtracted.
Although such events are produced at \(z \sim 1\) and the inelasticity
was restricted to \(0.2 < z < 0.9\), some diffractive events migrate into
the data sample due to the finite \(z\) resolution.
The reconstructed track multiplicity distribution was fitted to the sum of
inelastic ({\sc Epjpsi}) and diffractive ({\sc Epsoft}) MC
predictions. The fit yielded a contribution of \(6\pm1\%\) from proton 
dissociation for the whole sample.
The proton--dissociative contributions were subtracted bin--by--bin from
all measured cross sections according to the {\sc Epsoft} predictions 
normalised to the above fraction. 

\begin{figure}
\unitlength1cm  \begin{picture}(15.5,20.)
\includegraphics{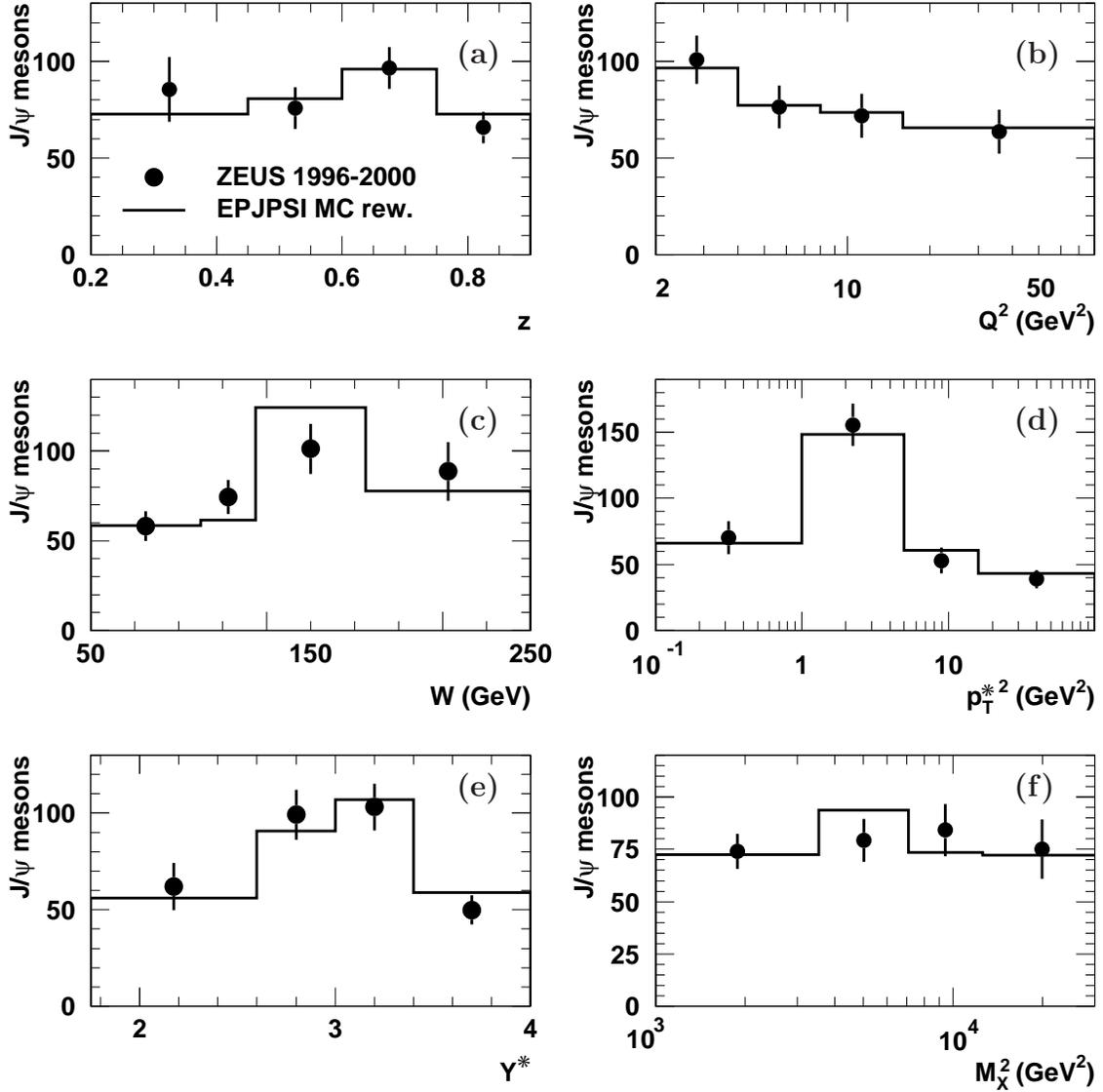}
\put(6.6,18.1){\textbf{(a)}}
\put(14.2,18.1){\textbf{(b)}}
\put(6.6,13.1){\textbf{(c)}}
\put(14.2,13.1){\textbf{(d)}}
\put(6.6,8.1){\textbf{(e)}}
\put(14.2,8.1){\textbf{(f)}}
\end{picture}
\vspace{-4.25cm}
\caption{
Number of \(J/\psi\) mesons reconstructed in the kinematic region
\(2<Q^2<80\gev^2\), \(50<W<250\gev\), \(0.2<z<0.9\) and
\(-1.6<Y_\text{lab}<1.3\) plotted as a function of (a) \(z\), 
(b) \(Q^2\), (c) \(W\), (d) \(p^{* 2}_T\), (e) \(Y^*\) and (f) \( M_X^2\).
The data distributions are shown as the points with statistical
errors only.
The histograms show the {\sc Epjpsi} MC predictions reweighted to the
data shapes in \(Q^2\) and \(p^{* 2}_T\) and area normalised to
the data.
}
\label{fig:cp}
\end{figure}

The number of \(J/\psi\) mesons reconstructed in the kinematic region
\(2<Q^2<80\gev^2\), \(50<W<250\gev\), \(0.2<z<0.9\) and
\(-1.6<Y_\text{lab}<1.3\), after subtraction of the 
proton--dissociative admixture, was compared to the predictions of the
{\sc Epjpsi} MC generator. The results are shown in Fig.~\ref{fig:cp}
for \(z\), \(Q^2\), \(W\), \(p^{* 2}_T\), the \(J/\psi\)
rapidity in the \(\gamma p\) frame\footnote{In the \(\gamma p\) 
centre--of--mass frame, the photon direction was chosen to be the
``forward direction''.}, \(Y^*\), and \( M_X^2\), where \(M_X\)
is the invariant mass of the final state excluding the \(J/\psi\) and
the scattered electron.

Data were corrected bin--by--bin for geometric acceptance, detector,
trigger and reconstruction inefficiencies, as well as for detector 
resolution, using the {\sc Epjpsi} MC generator.
The acceptance, $A_i({\mathcal O})$, as a function of an observable, ${\mathcal O}$, in a given
bin, $i$, is \mbox{$A_i({\mathcal O})=N_i^{\rm rec}({\mathcal O})/
N_i^{\rm gen}({\mathcal O})$}, where
$N_i^{\rm gen}({\mathcal O})$ is the number of generated MC
events and $N_i^{\rm rec}({\mathcal O})$ is the number of reconstructed
events passing all the selection requirements.

Differential cross sections as a function of ${\mathcal O}$
in a given bin $i$ were obtained using the expression
\[
\frac{d \sigma_i}{d{\mathcal O}} = 
\frac{N_i}
     {{\mathcal B}~{\mathcal L}~A_i({\mathcal O})},
\]
where $N_i$ is the number of signal events, reconstructed in each bin 
after subtraction of the estimated contribution from the diffractive 
proton--dissociative events, ${\mathcal B}$ the branching ratio (5.88 $\pm$ 0.10)\%
~\cite{pl:b592:1} and ${\mathcal L}$ the integrated luminosity.

The background from \(\psi'\) photoproduction
is expected to be \(15\%\)~\cite{pl:b348:657,*np:b459:3}; this
expectation was confirmed by a direct measurement of the
\(\psi'\) to \(J/\psi\) cross section ratio~\cite{epj:c27:173}.
Restricting the phase--space region in this analysis similar to that for 
photoproduction, \(50<W<180\gev\) and \(0.55<z<0.9\), 
the number of observed \(\psi'\) events was consistent with the
expectation from the \(\psi'\) to \(J/\psi\) ratio measured in the 
photoproduction regime.
The contribution of \(J/\psi\) mesons from \(\psi'\) decays
was assumed to yield
the same kinematic distributions as the dominant direct \(J/\psi\) 
contribution and, therefore, the theoretical predictions for 
\(J/\psi\) production were scaled up by 15\%. This change is small
compared to the normalisation error of the LO NRQCD predictions.

Monte Carlo studies showed that the contribution from
\(B\)--meson decays into \(J/\psi\) was concentrated at low--\(z\) values 
and small elsewhere.
For \(0.1 < z < 0.4 \), this contribution can be as large as 20\%.
The beauty contribution was estimated using the {\sc Rapgap} MC
and added to the \(J/\psi\) predictions.
This change is small compared to the normalisation uncertainty of the LO 
NRQCD predictions.

The \(J/\psi\) meson can be produced via \(\chi_c\) radiative decays, 
\(\chi_c \rightarrow J/\psi \gamma\).
While \(\chi_c\) mesons can be produced copiously in hadron--hadron 
collisions through \(g g\), \(g q\) and \(q \overline{q}\) interactions,
\(\chi_c\) production  via photon--gluon fusion is forbidden at LO
in the CS model.
This leaves only resolved photon processes,
strongly suppressed at non--zero photon virtuality, or CO processes 
as sources of \(\chi_c\) production.
However, the ratio of the \(\chi_c\) to \(J/\psi\) from
the CO processes is expected to be below 1\%~\cite{hep-ph-0106120}.
This contribution was therefore neglected.

The effect of the LO electroweak corrections was studied using the
{\sc Heracles}~\cite{cpc:69:155} MC program.
The open charm DIS cross section was evaluated using
the {\sc Rapgap}~\cite{cpc:86:147} MC program with and without radiative
corrections, as calculated by {\sc Heracles}, in a \(W\)--\(Q^2\) grid.
The measured cross sections were then corrected to the QED Born level
using the {\sc Heracles} predictions. In the region covered by the data, this
correction was $-2$\% on average and always below 7\% in absolute value.

\section{Systematic uncertainties}

The systematic uncertainties of the measured differential cross sections
were determined by changing the selection cuts or the analysis procedure
in turn and repeating the extraction of the differential cross sections.
The resulting uncertainty on the total cross section is given in 
parentheses.
The following categories of systematic uncertainties were considered:

\begin{itemize}
\item scattered electron reconstruction: these uncertainties were evaluated 
as described elsewhere~\cite{pr:d69:12004} (2\%);

\item CAL energy scale and resolution simulation: these uncertainties were 
evaluated as described elsewhere~\cite{pr:d69:12004} (2\%);

\item tracking: the resolutions on 
track momenta and angles were varied by $\pm$ 20\% of their values 
and the magnetic field by $\pm$ 0.3\% (1\%);

\item muon reconstruction: the uncertainty of the muon acceptance, including 
those of the efficiency of the muon chambers, the trigger
selection algorithms and the offline reconstruction, was obtained
from a study based on an independent dimuon sample at high \(Q^2\),
performed following the method discussed elsewhere~\cite{thesis:turcato:2002}
(6\%);

\item fitting procedure: the invariant--mass range and the functional form 
of the background were varied (2\%);

\item simulation of the process \(\gamma^{*} g \to J/\psi g\):
the {\sc Cascade} MC rather than the {\sc Epjpsi} MC was used to
calculate acceptances (5\%); 

\item subtraction of the remaining diffractive proton--dissociative
admixture: {\sc Diffvm} rather than {\sc Epsoft} was used to
perform the subtraction of the proton diffractive events (3\%).
\end{itemize}
These estimations were also made in each bin of the differential
cross sections.
All of the above individual sources of systematic uncertainty were added
in quadrature.

The following sources resulted in an overall shift of the cross section
and were therefore treated as normalisation uncertainties:
\begin{itemize}
\item the integrated luminosity determination has an uncertainty of
2\%;
\item the branching ratio of \( J/\psi \rightarrow \mu^+ \mu^- \) 
has an uncertainty of 1.7\%~\cite{pl:b592:1}.
\end{itemize}
The normalisation uncertainties were not included in the total 
systematic uncertainty.

\section{Results}
\label{sec-res}

\begin{figure}
\unitlength1cm  \begin{picture}(15.5,20.)
\includegraphics{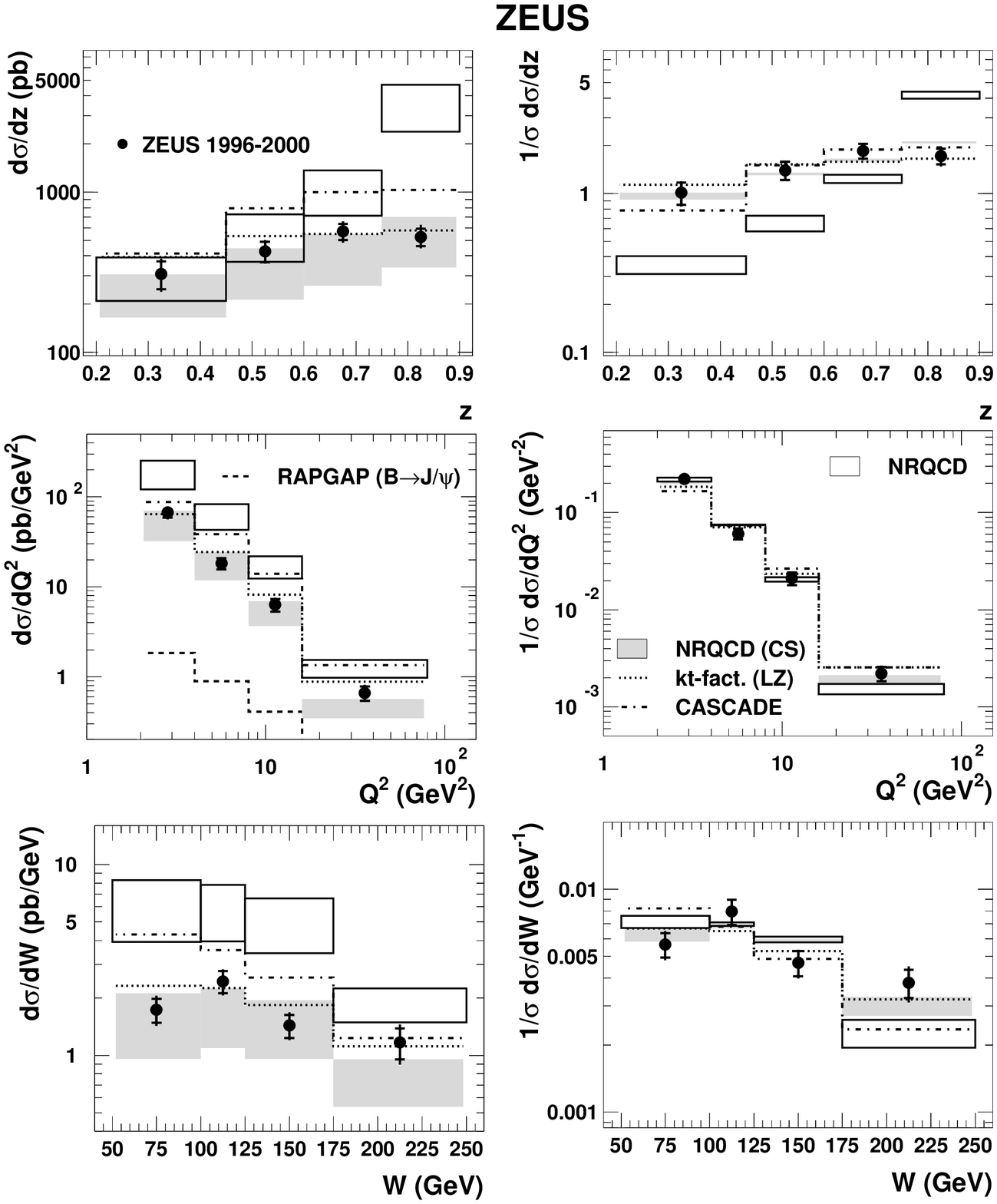}
\put(6.2,16.25){\textbf{(a)}}
\put(14.6,16.25){\textbf{(b)}}
\put(6.2,13.){\textbf{(c)}}
\put(14.6,13.){\textbf{(d)}}
\put(6.2,7.5){\textbf{(e)}}
\put(14.6,7.5){\textbf{(f)}}
\end{picture}
\vspace{-2.0cm}
\caption{
Differential cross sections for the reaction \(e\,p\to e\,J/\psi\,X\)
in the kinematic region \(2<Q^2<80\gev^2\), \(50<W<250\gev\),
\(0.2<z<0.9\) and \(-1.6<Y_\text{lab}<1.3\) as a function of (a) \(z\),
(c) \(Q^2\) and (e) \(W\).
The inner error bars of the data points show the statistical uncertainty;
the outer bars show statistical and systematic uncertainties added
in quadrature. 
The data are compared to LO NRQCD predictions, a LO CS calculation,
a prediction in the \(k_T\)--factorisation approach within the CSM 
and the {\sc Cascade} MC predictions.
(b), (d) and (f) show the data and the theoretical predictions
normalised to unit area.
}
\label{fig:zc1}
\end{figure}

\begin{figure}
\unitlength1cm  \begin{picture}(15.5,20.)
\includegraphics{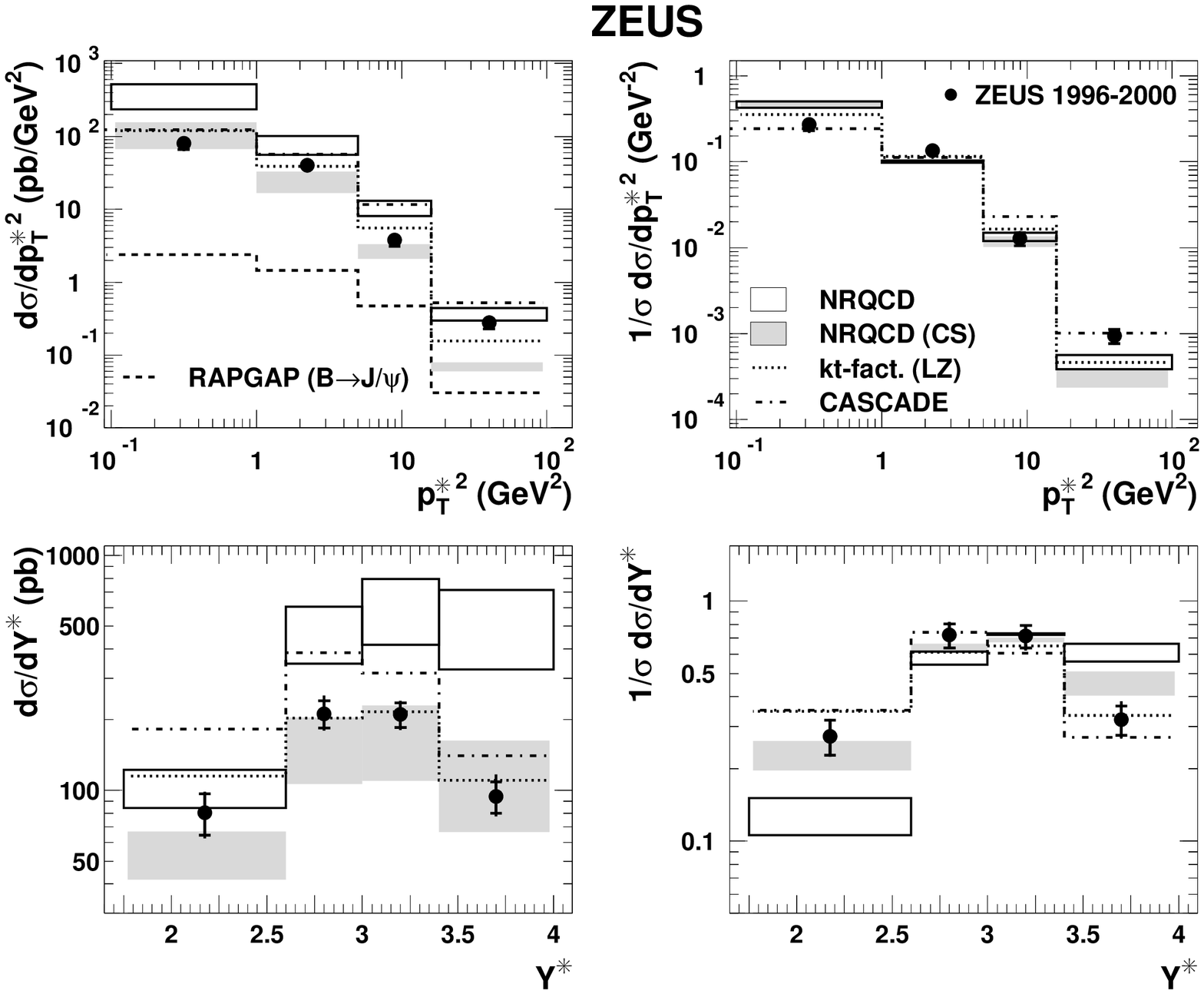}
\put(6.2,14.75){\textbf{(a)}}
\put(14.6,14.75){\textbf{(b)}}
\put(6.2,5.5){\textbf{(c)}}
\put(14.6,5.5){\textbf{(d)}}
\end{picture}
\vspace{-4.0cm}
\caption{
Differential cross sections for the reaction \(e\,p\to e\,J/\psi\,X\)
in the kinematic region \(2<Q^2<80\gev^2\), \(50<W<250\gev\),
\(0.2<z<0.9\) and \(-1.6<Y_\text{lab}<1.3\) as a function of 
(a) \(p^{* 2}_T\) and (c) \(Y^*\).
The inner error bars of the data points show the statistical uncertainty;
the outer bars show statistical and systematic uncertainties added
in quadrature. 
The data are compared to LO NRQCD predictions, a LO CS calculation,
a prediction in the \(k_T\)--factorisation approach within the CSM
and the {\sc Cascade} MC predictions.
(b) and (d) show the data and the theoretical predictions
normalised to unit area.
}
\label{fig:zc2}
\end{figure}

\begin{figure}
\unitlength1cm  \begin{picture}(15.5,20.)
\includegraphics{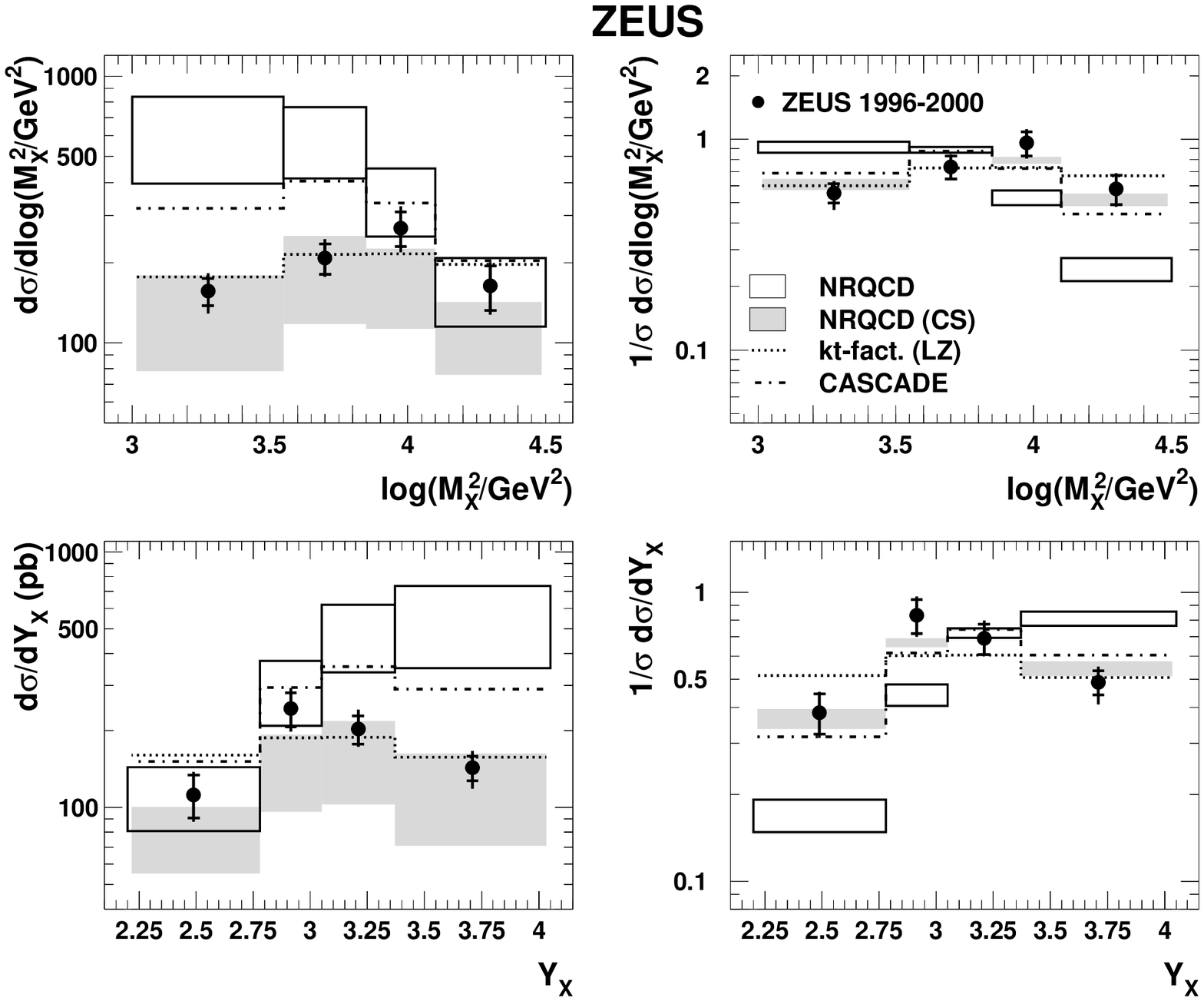}
\put(6.2,15.75){\textbf{(a)}}
\put(14.6,15.75){\textbf{(b)}}
\put(6.2,5.5){\textbf{(c)}}
\put(14.6,5.5){\textbf{(d)}}
\end{picture}
\vspace{-4.0cm}
\caption{
Differential cross sections for the reaction \(e\,p\to e\,J/\psi\,X\)
in the kinematic region \(2<Q^2<80\gev^2\), \(50<W<250\gev\),
\(0.2<z<0.9\) and \(-1.6<Y_\text{lab}<1.3\) as a function of
(a) \( \log M_X^2\) and (c) \( Y_X \).
The inner error bars of the data points show the statistical uncertainty;
the outer bars show statistical and systematic uncertainties added
in quadrature. 
The data are compared to LO NRQCD predictions, a LO CS calculation,
a prediction in the \(k_T\)--factorisation approach within the CSM
and the {\sc Cascade} MC predictions.
(b) and (d) show the data and the theoretical predictions
normalised to unit area.
}
\label{fig:zc3}
\end{figure}

The cross section for the process \( e p \rightarrow e J/\psi X \)
in the kinematic region \(2<Q^2<80\gev^2\), \(50<W<250\gev\), 
\(0.2<z<0.9\) and \(-1.6<Y_\text{lab}<1.3\) is
\begin{center}
302 $\pm$ 23 ({\rm stat.}) $^{+28}_{-20}$ ({\rm syst.}) {\rm pb},
\end{center}
where the first uncertainty is statistical and the second systematic.
In Figs.~\ref{fig:zc1}, \ref{fig:zc2} and \ref{fig:zc3}, the differential 
cross sections as a function of \(z\), \(Q^2\), \(W\), \(p^{* 2}_T\),
\(Y^*\), \( \log M_X^2\) and the rapidity of the hadronic system \(X\),
\( Y_X \), are shown.
They are compared to the predictions of a NRQCD model~\cite{np:b621:337},
a CS model with \(k_T\) factorisation (LZ)~\cite{epj:c27:87} and to 
the {\sc Cascade} MC.
The beauty contribution, estimated using the {\sc Rapgap} MC, is also shown
separately in Figs.~\ref{fig:zc1}c and \ref{fig:zc2}a.
All differential cross sections and normalised cross sections are given 
in Tables~\ref{tab:zeusps1}~and~\ref{tab:zeusps2}.

The uncertainties
for the CS and CO NRQCD predictions correspond to variations of
the charm--quark mass
(\(m_c=1.5\pm0.1\gev\)) and of the renormalisation and factorisation scales 
from \(\by12\sqrt{\smash[b]{Q^2+M^2_\psi}}\) to 
\(2\sqrt{\smash[b]{Q^2+M^2_\psi}}\).
The uncertainty on the long--distance matrix elements and the effect 
of different choices of parton distribution functions (default set is MRST98LO)
are also taken into account.
The bands in the figures shows all these uncertainties added in quadrature.

In general, the CSM is consistent with the data.
The predictions including both CS and CO contributions are higher than
the data, especially at high \(z\) and low \(p_T^{* 2}\). 
At high values of \(p^{* 2}_T\) the agreement with the data is 
reasonable.
The prediction does not describe the shapes of the \(z\), \(Y^*\),
\( \log M_X^2\) and \( Y_X \) distributions.
Previous photoproduction results~\cite{epj:c25:25,epj:c27:173} showed that
the agreement between data and theory at high \(z\) can be improved
using resummed LO NRQCD predictions~\cite{pr:d62:34004}.
It should be noted that, in photoproduction, inclusion of the NLO
corrections to the CSM, not available for DIS, significantly improved
the description of the data.

For the LZ \(k_T\)--factorisation predictions,
the parametrisation, KMS~\cite{pr:d56:3991}, of the unintegrated gluon density 
was used.
The charm--quark mass was set to \(m_c=1.4\gev\), which is the mass used
in the KMS parametrisation. The renormalisation and factorisation scales
were both set to \(\mu=k_T\) for \(k_T > 1 \gev\). For \(k_T~\leq~1 \, \gev\)
the scales were fixed at \(1 \gev\).
Calculations based on the \(k_T\)-factorisation approach give a reasonable 
description of the data both in shape and normalisation. 

The data are also compared with the predictions of the \textsc{Cascade}
MC using the $k_T$-factorisation approach, where gluons are treated 
according to the CCFM evolution equations.
These predictions were obtained by setting the charm--quark
mass to 1.5 \(\gev\), the evolution scale of the strong coupling constant
to the \(J/\psi\) transverse mass, \( \sqrt{M^2_\psi+p_T^2} \),
and using the unintegrated gluon--density parametrisation ``J2003 set 2''.
The \textsc{Cascade} MC is above the data for \(z > 0.45\)
and for \(W < 175 \gev\).

\begin{figure}
\unitlength1cm  \begin{picture}(15.5,20.)
\includegraphics{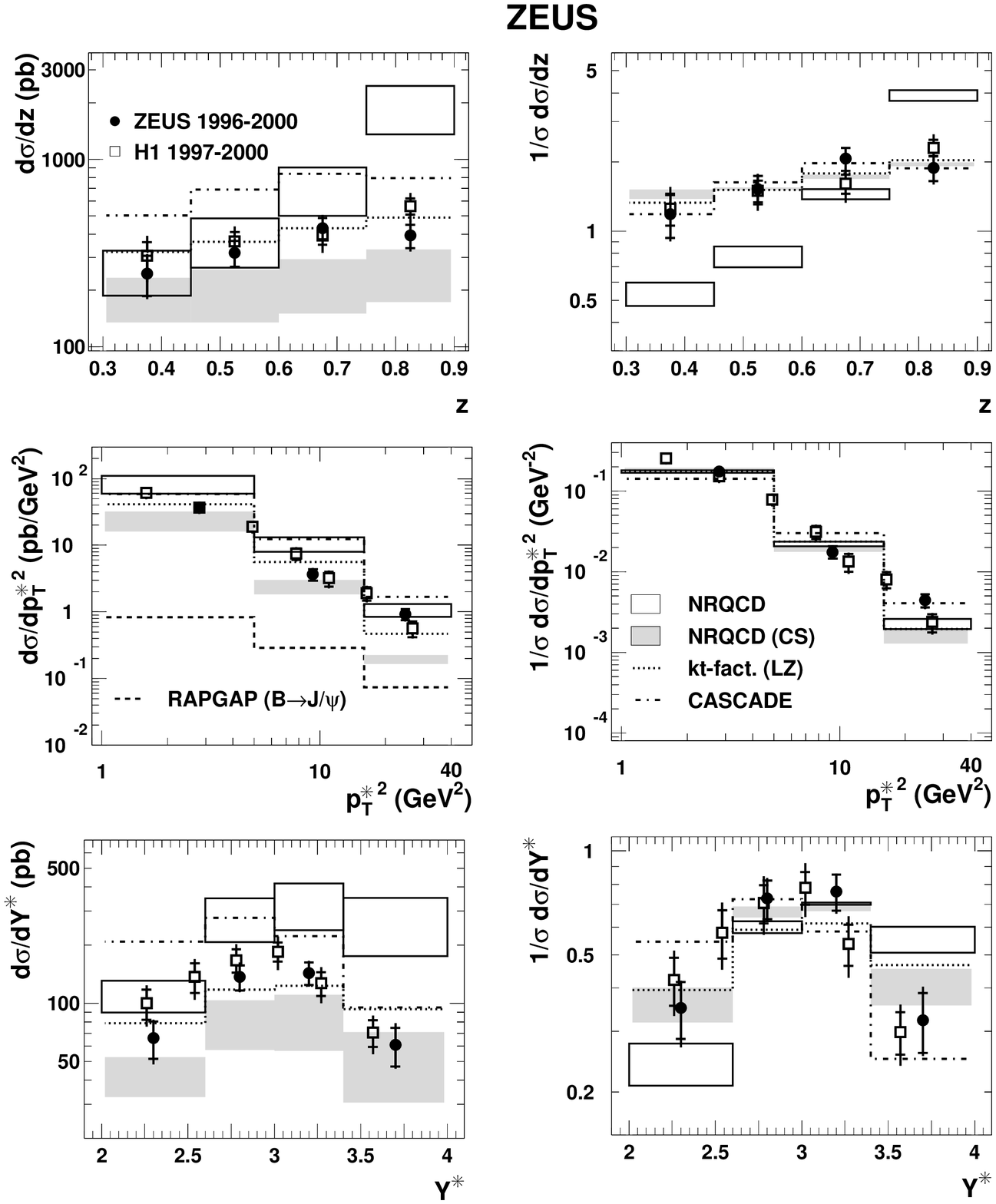}
\put(6.2,17.0){\textbf{(a)}}
\put(14.6,17.0){\textbf{(b)}}
\put(6.2,14.4){\textbf{(c)}}
\put(14.6,14.4){\textbf{(d)}}
\put(1.2,7.9){\textbf{(e)}}
\put(9.8,7.9){\textbf{(f)}}
\end{picture}
\vspace{-2.75cm}
\caption{
Differential cross sections for the reaction \(e\,p\to e\,J/\psi\,X\)
in the kinematic region \(2<Q^2<100\gev^2\), \(50<W<225\gev\), \(0.3<z<0.9\) 
and \(p_T^{*2}>1\gev^2\) as a function of (a) \(z\), (c) \(p^{* 2}_T\) 
and (e) \(Y^*\).
The inner error bars of the data points show the statistical uncertainty;
the outer bars show statistical and systematic uncertainties added
in quadrature.
The ZEUS and H1 data are compared to LO NRQCD predictions, a LO CS calculation,
a prediction in the \(k_T\)--factorisation approach within the CSM
and the {\sc Cascade} MC predictions.
The H1 data points are plotted at the mean value of the data in each 
interval~\protect\cite{epj:c25:41}. 
The ZEUS data for the \(p^{* 2}_T\) differential cross section are plotted at 
the weighted mean, for each bin, of the {\sc Epjpsi} MC prediction.
(b), (d) and (f) show the data and the theoretical predictions
normalised to unit area.
}
\label{fig:zvh}
\end{figure}

In order to compare the present measurements directly to the H1 
results~\cite{epj:c25:41}, differential cross sections were determined in the kinematic
range \(2<Q^2<100\gev^2\), \(50<W<225\gev\), \(0.3<z<0.9\) and 
\(p_T^{*2}>1\gev^2\);
all ZEUS differential cross sections and normalised cross sections are
given in Table~\ref{tab:h1ps}.
The results of this comparison are shown in Fig.~\ref{fig:zvh}.
The present results are in agreement with those from H1.
In Fig.~\ref{fig:zc1}a, the ZEUS data are in better agreement with 
the CSM prediction than in Fig.~\ref{fig:zvh}a. This is a consequence
of the \(p_T^{*2}>1\gev^2\) cut used in Fig.~\ref{fig:zvh}a combined with 
the fact that the CS prediction underestimate the
data at high \(p_T^{*2}\), as seen in Fig.~\ref{fig:zc2}a.

\section{Conclusions}
\label{sec-sum}

Inelastic \(J/\psi\) production in DIS has been measured in the
kinematic region \(2<Q^2<80\gev^2\), \(50<W<250\gev\), \(0.2<z<0.9\)
and \(-1.6<Y_\text{lab}<1.3\).
The data are in agreement with the H1 results in the kinematic region
\(2<Q^2<100\gev^2\), \mbox{\(50<W<225\gev\)}, \(0.3<z<0.9\) and 
\(p_T^{*2}>1\gev^2\).
The data are compared with LO NRQCD predictions, including both CS and CO 
contributions, and \(k_T\)--factorisation calculations.
Calculations of the CS process generally agree with the
data, whereas inclusion of CO terms spoils this agreement.

\section*{Acknowledgments}
\vspace{0.3cm}
We thank the DESY Directorate for their strong support and encouragement.
The remarkable achievements of the HERA machine group were essential for
the successful completion of this work.
The design, construction and installation of the ZEUS detector have been
made possible by the effort of many people who are not listed as authors.
It is a pleasure to thank B.A. Kniehl, A.V. Lipatov, C.P. Palisoc,
N.P. Zotov and L. Zwirner for providing us with their calculations.
We are grateful to H. Jung for helpful discussions.


{
\def\bibname{\Large\bf References}
\def\refname{\Large\bf References}
\pagestyle{plain}
\ifzeusbst
  \bibliographystyle{./BiBTeX/bst/l4z_default}
\fi
\ifzdrftbst
  \bibliographystyle{./BiBTeX/bst/l4z_draft}
\fi
\ifzbstepj
  \bibliographystyle{./BiBTeX/bst/l4z_epj}
\fi
\ifzbstnp
  \bibliographystyle{./BiBTeX/bst/l4z_np}
\fi
\ifzbstpl
  \bibliographystyle{./BiBTeX/bst/l4z_pl}
\fi
{\raggedright
\bibliography{./BiBTeX/user/syn.bib,%
              ./BiBTeX/bib/l4z_articles.bib,%
              ./BiBTeX/bib/l4z_books.bib,%
              ./BiBTeX/bib/l4z_conferences.bib,%
              ./BiBTeX/bib/l4z_h1.bib,%
              ./BiBTeX/bib/l4z_misc.bib,%
              ./BiBTeX/bib/l4z_old.bib,%
              ./BiBTeX/bib/l4z_preprints.bib,%
              ./BiBTeX/bib/l4z_replaced.bib,%
              ./BiBTeX/bib/l4z_temporary.bib,%
              ./BiBTeX/bib/l4z_zeus.bib,%
              ./BiBTeX/bib/l4z_zzzmy.bib}}
}
\vfill\eject

%
%
%
\begin{table}[p]
\begin{center}
\begin{tabular}{||c|c|c||} 
\hline
\(z\) range   & \(d\sigma/dz\) (pb) & \(1/\sigma d\sigma/dz\) \\ 
\hline\hline
\(0.20\rnge0.45\) & \(309\pm61^{+41}_{-34}\) & \(1.01\pm0.16^{+0.09}_{-0.08}\) \\
\(0.45\rnge0.60\) & \(428\pm62^{+44}_{-32}\) & \(1.{40}\pm0.19^{+0.09}_{-0.06}\) \\
\(0.60\rnge0.75\) & \(568\pm65^{+64}_{-55}\) & \(1.86\pm0.{20}^{+0.08}_{-0.13}\) \\
\(0.75\rnge0.90\) & \(526\pm66^{+74}_{-47}\) & \(1.72\pm0.{20}^{+0.17}_{-0.15}\) \\
\hline\hline
\(W\) range (\Gev) &  \(d\sigma/dW\) (\(\text{pb}/\Gev\)) & \(1/\sigma d\sigma/dW\) \\
\hline\hline
\(50\rnge100\)  & \(1.73\pm0.25^{+0.20}_{-0.16}\) & \(0.0056\pm0.0007^{+0.0005}_{-0.0005}\) \\
\(100\rnge125\) & \(2.44\pm0.32^{+0.23}_{-0.20}\) & \(0.{0080}\pm0.{0010}^{+0.0004}_{-0.0005}\) \\
\(125\rnge175\) & \(1.43\pm0.20^{+0.14}_{-0.12}\) & \(0.0047\pm0.0006^{+0.0003}_{-0.0003}\) \\
\(175\rnge250\) & \(1.17\pm0.22^{+0.19}_{-0.17}\) & \(0.0038\pm0.0006^{+0.0004}_{-0.0004}\) \\
\hline\hline
\(Q^2\) range (\(\Gev^2\)) & \(d\sigma/dQ^2\) (\(\text{pb}/\Gev^2\)) & \(1/\sigma d\sigma/dQ^2\) \\
\hline\hline
\(2\rnge4\)   & \(66.9\pm8.4^{+7.7}_{-6.8}\)    & \(0.{223}\pm0.019^{+0.008}_{-0.012}\)  \\
\(4\rnge8\)   & \(18.3\pm2.7^{+1.6}_{-1.3}\)    & \(0.0609\pm0.0079^{+0.0033}_{-0.0028}\)  \\
\(8\rnge16\)  & \(6.3\pm1.{0}^{+0.7}_{-0.6}\)   & \(0.0211\pm0.0032^{+0.0013}_{-0.0015}\)  \\
\(16\rnge80\) & \(0.66\pm0.12^{+0.09}_{-0.05}\) & \(0.00221\pm0.00038^{+0.00026}_{-0.00014}\)  \\
\hline\hline
\(p_T^{*2}\) range  (\(\Gev^2\)) & \(d\sigma/dp_T^{*2}\) (\(\text{pb}/\Gev^2\)) & \(1/\sigma d\sigma/dp_T^{*2}\) \\
\hline\hline
\(0\rnge1\)    & \(80\pm14^{+8}_{-9}\)              & \(0.269\pm0.041^{+0.012}_{-0.034}\)  \\
\(1\rnge5\)    & \(40.1\pm4.1^{+5.7}_{-2.6}\)       & \(0.1345\pm0.0096^{+0.{0080}}_{-0.0014}\)  \\
\(5\rnge16\)   & \(3.81\pm0.{70}^{+0.44}_{-0.32}\)  & \(0.0128\pm0.0022^{+0.{0010}}_{-0.0008}\)  \\
\(16\rnge100\) & \(0.{280}\pm0.051^{+0.031}_{-0.027}\) & \(0.00094\pm0.00017^{+0.00006}_{-0.00009}\)  \\
\hline\hline
\(Y^{*}\) range   &  \(d\sigma/dY^{*}\) (pb) & \(1/\sigma d\sigma/dY^{*}\) \\
\hline\hline
\(1.75\rnge2.60\) & \(80\pm16^{+9}_{-7}\)    & \(0.274\pm0.045^{+0.017}_{-0.021}\) \\
\(2.60\rnge3.00\) & \(212\pm28^{+30}_{-16}\) & \(0.722\pm0.083^{+0.052}_{-0.031}\) \\
\(3.00\rnge3.40\) & \(211\pm25^{+18}_{-16}\) & \(0.716\pm0.077^{+0.026}_{-0.055}\) \\
\(3.40\rnge4.00\) & \(94\pm14^{+18}_{-9}\)   & \(0.321\pm0.045^{+0.041}_{-0.024}\) \\
\hline
\end{tabular} 
\caption{Differential cross sections and normalised 
differential cross sections in the kinematic region
\(2<Q^2<80\gev^2\), \(50<W<250\gev\), \(0.2<z<0.9\) and 
\(-1.6<Y_\text{lab}<1.3\) as a function of \(z\), \(W\),
\(Q^2\), \(p_T^{*2}\) and \(Y^{*}\).
The first uncertainty is statistical and the second is
systematic. Overall normalisation uncertainties due to
the luminosity measurement (\(\pm2\%\)) and to the \(J/\psi\)
decay branching ratio (\(1.7\%\)) are not included in the
systematic error.}
\label{tab:zeusps1}
\end{center}
\end{table}
\begin{table}[p]
\begin{center}
\begin{tabular}{||c|c|c||}
\hline
\(\log(M_X^2/\Gev^2)\) range & \(d\sigma/d\log(M_X^2/\Gev^2)\) (pb) & 
\(1/\sigma d\sigma/d\log(M_X^2/\Gev^2)\) \\
\hline\hline
\(3.00\rnge3.55\) & \(156\pm18^{+19}_{-20}\)  & \(0.556\pm0.057^{+0.052}_{-0.074}\) \\
\(3.55\rnge3.85\) & \(208\pm27^{+25}_{-16}\)  & \(0.{740}\pm0.091^{+0.056}_{-0.021}\) \\
\(3.85\rnge4.10\) & \(270\pm40^{+38}_{-31}\)  & \(0.96\pm0.13^{+0.09}_{-0.08}\) \\
\(4.10\rnge4.50\) & \(164\pm31^{+21}_{-18}\)  & \(0.581\pm0.092^{+0.054}_{-0.{050}}\) \\
\hline\hline
\(Y_X\) range   & \(d\sigma/dY_X\) (pb) & \(1/\sigma d\sigma/dY_X\) \\
\hline\hline
\(2.20\rnge2.78\) & \(112\pm21^{+13}_{-11}\)  & \(0.383\pm0.061^{+0.033}_{-0.033}\) \\
\(2.78\rnge3.05\) & \(243\pm37^{+33}_{-26}\)  & \(0.83\pm0.11^{+0.08}_{-0.07}\) \\
\(3.05\rnge3.37\) & \(203\pm26^{+29}_{-15}\)  & \(0.692\pm0.083^{+0.067}_{-0.018}\) \\
\(3.37\rnge4.05\) & \(143\pm16^{+17}_{-18}\)  & \(0.488\pm0.047^{+0.044}_{-0.063}\) \\
\hline
\end{tabular}
\caption{Differential cross sections and normalised 
differential cross sections in the kinematic region
\(2<Q^2<80\gev^2\), \(50<W<250\gev\), \(0.2<z<0.9\) and 
\(-1.6<Y_\text{lab}<1.3\) as a function of \(\log(M_X^2/\Gev^2)\) 
and \(Y_X\).
The first uncertainty is statistical and the second is
systematic. Overall normalisation uncertainties due to
the luminosity measurement (\(\pm2\%\)) and to the \(J/\psi\)
decay branching ratio (\(1.7\%\)) are not included in the
systematic error.}
\label{tab:zeusps2}
\end{center}
\end{table}
%
%
\begin{table}[p]
\begin{center}
\begin{tabular}{||c|c|c||} 
\hline
\(z\) range & \(d\sigma/dz\) (pb) & \(1/\sigma d\sigma/dz\) \\
\hline
\hline
\(0.30\rnge0.45\) & \(246\pm60^{+28}_{-29}\) & \(1.18\pm0.25^{+0.07}_{-0.13}\) \\
\(0.45\rnge0.60\) & \(317\pm50^{+39}_{-24}\) & \(1.53\pm0.22^{+0.13}_{-0.{10}}\) \\
\(0.60\rnge0.75\) & \(430\pm56^{+51}_{-34}\) & \(2.07\pm0.23^{+0.09}_{-0.11}\) \\
\(0.75\rnge0.90\) & \(392\pm57^{+64}_{-41}\) & \(1.89\pm0.24^{+0.22}_{-0.17}\) \\
\hline\hline
\(p_T^{*2}\) range  (\(\Gev^2\)) & \(d\sigma/dp_T^{*2}\) (\(\text{pb}/\Gev^2\)) & \(1/\sigma d\sigma/dp_T^{*2}\) \\
\hline\hline
\(1\rnge5\)    & \(36.4\pm3.7^{+4.1}_{-2.4}\)       & \(0.1752\pm0.0092^{+0.0054}_{-0.0054}\) \\
\(5\rnge16\)   & \(3.65\pm0.71^{+0.17}_{-0.32}\)    & \(0.0176\pm0.{0030}^{+0.0022}_{-0.0011}\) \\
\(16\rnge40\)  & \(0.92\pm0.18^{+0.12}_{-0.{10}}\)  & \(0.00443\pm0.00083^{+0.00028}_{-0.00051}\) \\
\hline
\hline
\(Y^{*}\) range   &  \(d\sigma/dY^{*}\) (pb)         & \(1/\sigma d\sigma/dY^{*}\) \\
\hline\hline
\(2.00\rnge2.60\) & \(66\pm14^{+7}_{-10}\)   & \(0.351\pm0.066^{+0.023}_{-0.048}\) \\
\(2.60\rnge3.00\) & \(137\pm20^{+15}_{-10}\) & \(0.728\pm0.094^{+0.053}_{-0.048}\) \\
\(3.00\rnge3.40\) & \(144\pm19^{+13}_{-11}\) & \(0.762\pm0.091^{+0.035}_{-0.049}\) \\
\(3.40\rnge4.00\) & \(61\pm14^{+12}_{-6}\)   & \(0.323\pm0.064^{+0.049}_{-0.025}\) \\
\hline
\end{tabular}
\caption{Differential cross sections and normalised 
differential cross sections in the kinematic region
\(2<Q^2<100\gev^2\), \(50<W<225\gev\), \(0.3<z<0.9\) and 
\(p_T^{*2}>1\gev^2\) as a function of \(z\), \(p_T^{*2}\) and \(Y^{*}\).
The first uncertainty is statistical and the second is
systematic. Overall normalisation uncertainties due to
the luminosity measurement (\(\pm2\%\)) and to the \(J/\psi\)
decay branching ratio (\(1.7\%\)) are not included in the
systematic error.}
\label{tab:h1ps}
\end{center}
\end{table}
%

%
%

%
%
\end{document}